\documentstyle[12pt]{article}
\newcommand{\be}{\begin{equation}}
\newcommand{\ee}{\end{equation}}
\newcommand{\bea}{\begin{eqnarray}}
\newcommand{\eea}{\end{eqnarray}}
\newcommand{\bd}{\begin{displaymath}}
\newcommand{\ed}{\end{displaymath}}

\begin{document}
\bibliographystyle{physics}
\renewcommand{\thefootnote}{\fnsymbol{footnote}}

\author{
Hongying Jin and Xinmin Zhang
\thanks{Email: jhy@hptc5.ihep.ac.cn,~~xmzhang@bepc3.ihep.ac.cn}\\
{\small\sl  Institute of High Energy Physics, Academia Sinica,
P.O.Box 918(4), Beijing 100039, China }\\
 }

\date{}
\title{
\vspace{3cm}
\bigskip
\bigskip
{\LARGE\sf Scalar Glueball Decay Into Pions In Effective Theory}
 } 
\maketitle
\begin{abstract}
\noindent

\end{abstract}
\vspace{1.5cm}

We examine the mixing between the sigma meson $\sigma$
and the "pure" glueball field $H$ and study the decays of the scalar glueball
candidate $f_0(1500)$ (a linear combination of $\sigma$ and $H$) into
two and four pions in effective linear sigma model. From
 recent experimental data on $f_0(1500)$ decay to pions we point out 
that the mixing angle is of order of 0.1.

\vspace{1.5cm}
{\bf PACS  numbers 13.25.Hw 12.28.Lg } 
\newpage

Quantum Chromo Dynamics (QCD), as
the fundamental theory of the strong
interaction predicts the existence of exotic mesons 
made of gluons. Observations 
of these gluonium will provide a 
direct confirmation on the special feature of non-abelian gauge theory.
For scalar glueballs
the candidates reported in the literature include $f_0(1500)$ and
$f_0(1750)$[1,2,3]. However, the non-perturbative dynamics of QCD
makes it
difficult to do the calculation on the glueball mass, decay width and mixing
with the sigma meson. Besides Lattice QCD\cite{latt}, one has considered
various phenomenological models, such as the potential models\cite{vo},
bag models\cite{bag}, flux tube\cite{flux}
models and the QCD sum rule\cite{sum} in order to 
study the glueball mass and its
two body decays. In this paper we take an effective lagrangian approach and 
discuss the glueball decay into two as well as four-pion final states.
Particularly we will pay attention to the mixing between the glueball and the
sigma meson.

It is well known that non-linear sigma model describes
the dynamical properties of the pions. To construct an effective
lagrangian for glueball decay into pions 
one could simply include the glueball meson
into the non-linear sigma model. Under the chiral symmetry,
the glueball transforms as a singlet. For instance, interaction
 responsible for the decay of the scalar glueball
into multi-pions is given by
\begin{equation}
H Tr{\partial}_\mu \Sigma^+ \partial^\mu \Sigma
\end{equation}
where $\Sigma=\exp{i {\vec \tau}\cdot {\vec \pi} / f_\pi}$ and $H$ the 
"pure" glueball field.  
However, with mass of scalar glueball around 1500 MeV or 1700MeV, the
momentum
transfer from the decay of the glueball into pions is so large that
resonances,
such as sigma and rho mesons will show up and play an essential role in
the processes of the glueball decay. 
Thus we use in this paper  an 
 effective
lagrangian
with sigma and rho mesons included explicitly {\it i.e.} the linear sigma
model.

The identification of the sigma particle $\sigma$ is highly controversial.
It is listed in the Particle Data Tables\cite{data} as a very broad meson
with mass
around $400 - 1200$ MeV and full width $600-1200$ MeV. This particle,
needed in the linear sigma model, has been argued to play an important
role
in nuclear physics\cite{ta} and in the study on the chiral phase
transition\cite{pisarski}. In 
recent years, many works are devoted to search for it in the $\pi \pi$ 
scattering\cite{s1}. Regarding its role in the
glueball physics, 
BES\cite{BES} and MKIII\cite{MKIII} data on
$B(J/\Psi\rightarrow\gamma f_0(1500)\rightarrow 
\gamma 4\pi) $ and 
$B(J/\Psi\rightarrow\gamma f_0(1750)\rightarrow 
\gamma 4\pi)$  
, Crystal Barrel data\cite{Bar} on  
${\bar p}p\rightarrow \pi^0 f_0(1500)\rightarrow \pi^0
(2\pi,{\bar K}K,2\eta,\eta'\eta,4\pi) $
 show $f_0(1500)$ decays into $4\pi$ dominantly through 
sigma channel(S partial wave), and also
 hints it strongly couples to 
$2\sigma$.



In this paper, we consider, for simplicity, the linear sigma model for two
flavors.
 As usual we introduce 
the field
\begin{equation}
\Phi=\displaystyle{\sigma\frac{\tau^0}{2}+i{\vec \pi}\frac{\vec \tau}{2}},
\end{equation} 
where $\tau^0$ is unity matrix and $\vec \tau$ the Pauli matrices with
normalization condition
  $Tr{\tau^a\tau^b}=2\delta^{ab}$.
Under a $SU_L(2)\otimes SU_R(2)$ chiral transformation, $\Phi$ 
transforms as 
\begin{equation}
\Phi\rightarrow L^\dagger\Phi R . 
\end{equation}
A renormalizable lagrangian of linear sigma model is given now by
\begin{eqnarray}
L_{\Phi-\Phi}=\displaystyle{Tr\{\partial_\mu\Phi^\dagger\partial^\mu\Phi\}
-\lambda[Tr\{\Phi^\dagger\Phi\}-\frac{f_\pi^2}{2}]^2},
\label{lal}
\end{eqnarray}
where $f_\pi$ is the vacuum expectation value of the
sigma field and $\lambda$ the self coupling 
constant.




Let's now add the $O^{++}$  "pure" glueball state H
on the lagrangian. 
Given that the
H is made of gluons and singlet under the chiral symmetry, there are
only two terms which give rise to the interaction among H, 
the sigma and pions,
\begin{eqnarray}
L_{H-\Phi}=\displaystyle{g_1 f_\pi H[Tr\{\Phi^\dagger\Phi\}
-\frac{f_\pi^2}{2}]+g_2 H^2[Tr\{\Phi^\dagger\Phi\}
-\frac{f_\pi^2}{2}]},
\label{laH}
\end{eqnarray}
where two free parameters, $g_1$ and
$g_2$ are introduced 
to describe the strength of couplings of one and two H to the
sigma or pions.

 
For the self interaction of glueballs, the lagrangian is given by
\begin{eqnarray}
L_{H-H}=\displaystyle{\frac{1}{2}\partial_\mu H\partial^\mu H+
\frac{1}{2}m_H^2H^2+f_3H^3+f_4H^4},
\end{eqnarray}
  where $m_H^2$ is the mass of glueball, $f_3, f_4$ the self coupling constants.

After the chiral symmetry is broken by non-vanishing vacuum expectation
value of the sigma field, lagrangian in (\ref{laH}) generates not only
the interaction
for glueball decay, but also a mixing between the glueball
 H and the sigma field
$\sigma$. The mass matrix 
for $(H, \sigma)$ is 
\begin{equation}
m^2=
\left(
\begin{array}{cc}
m_H^2 & g_1f^2_\pi\\
g_1f^2_\pi & m^2_\sigma
\end{array}
\right ) ,
\end{equation}
where $m_\sigma^2=2\lambda f^2_\pi$.

To diagonalize the mass matrix,
we introduce an unitary matrix 
\begin{equation}
V=
\left(
\begin{array}{cc}
cos\theta & sin\theta\\
-sin\theta & cos\theta
\end{array}
\right )   
\sim\left(
\begin{array}{cc}
1 & \theta\\ 
-\theta & 1  
\end{array}
\right ) .
\end{equation}
 Then the mass eigen-states are given by 
\begin{equation}
\left(
\begin{array}{c}      
H'\\
\sigma'
\end{array}
\right )
=
\left(
\begin{array}{cc}      
1 & \theta\\
-\theta & 1
\end{array}
\right )
\left(
\begin{array}{c}       
H\\
\sigma
\end{array}
\right ) ,
\label{mass}
\end{equation}
with $\theta= g_1 f_\pi^2 / (m_H^2 - m_\sigma^2)$.

Substituting (\ref{mass}) into (\ref{laH}) and (4), we obtain the
lagrangian
 for $H'$ decay
\begin{equation}
L_{H'}=\displaystyle{H'\theta\{\frac{m_{H'}^2}{2 f_\pi}{\bf
\pi}^2+(\frac{m_{H'}^2+2m^2_{\sigma'}}
{2 f_\pi }- 2 g_2 f_\pi ){\sigma'}^2 +O(\theta^2)\}} .
\label{decay}
\end{equation} 

In (\ref{decay}),
  the couplings of the glueball to sigma and the pions
 are proportional to the
mixing angle $\theta$. This could be understood by the observation that
in the limit of large $N_c$,
the mixing between $H$ and $\sigma$ 
is of the same order as that for 
 $H'$ to decay into light hadrons (see figure 1).
Furthermore,
under the assumption of vector meson dominance\cite{pisarski},
 $\rho$ meson can be introduced
by
 replacing $\partial^\mu\Phi$ in (\ref{lal})
by $D^\mu\Phi=\partial^\mu\Phi-ig(\rho^\mu\Phi-\Phi \rho^\mu)$. We found
that
the direct coupling of
 $H'$ 
and $\rho\rho$ vanishes at the tree level.

Identifying the glueball state
$H^\prime$ with $f_0(1500)$, the phenomenologies of our model can be
summarized as follows: 

i) Glueball decay into two pions: The decay width of $f_0(1500)$ into two
pions is given by 
\begin{equation}
\Gamma ( H^\prime \rightarrow 2\pi )
= \frac{3\theta^2}{32 \pi}
(\frac{m_{H'}}{f_\pi})^2
\sqrt{m_{H^\prime}^2 -4 m_\pi^2}.
\end{equation}

Experimentally Crystal Barrel Group reported\cite{Bar} that 
$Br(f_0(1500)\rightarrow 2\pi):Br(f_0(1500)\rightarrow 4\pi)=
4.39\pm0.16:14.9\pm3.2$ and
 $\Gamma_{total}=120\pm20MeV$.
 Assuming that $Br( f_0(1500)\rightarrow 4\pi )$
is about $ 50\%$ one has that $\Gamma(2\pi)=17.5\pm 6\pm 3 MeV$. 
Using this value, we obtain
\begin{equation} 
\theta\sim 0.05.
\label{ss}
\end{equation}



ii)Glueball decay into four pions: the dominant decay of the glueball into
four pions is through the intermediate two sigma states.
In Figure 2, we
plot the ratio of glueball decay into  $2\sigma$ to $2\pi$ as
function
of $g_2$. This measurement can be used to determine the parameter
$g_2$.

  

Before concluding our paper, we make two remarks:

1) In the chiral limit which we work on in 
this paper, the mixing of glueball with
sigma is a consequence of the spontaneous chiral symmetry breaking.
However, when including the explicit symmetry breaking effects by the quark
mass, especially
when extending the lagrangian for two flavors to three flavors, there would be
one term proportional to the explicit symmetry breaking which also generates
mixing of the glueball with the sigma meson as shown in Fig. 3;

2) To make a comparison of our result from effective theory with other
models,
we estimate the mixing angle $\theta$, as an example, 
by using QCD sum rule . 
The low energy theorem 
states that \cite{shif}
\begin{equation}
i\int dx<|TO(x){\cal O}|>=-d_O<|O(0)|>,
\end{equation}
where $d_O$ is the mass dimension of operator $O$, 
${\cal O}$
is the trace of the energy momentum tensor 
$$\theta^\mu_\mu=\displaystyle{\frac{\beta(\alpha_s)}{4\alpha_s}
G^{\mu\nu}G_{\mu\nu} \approx
\frac{-b\alpha_s}{8\pi}G^{\mu\nu}G_{\mu\nu}},$$
($\beta(\alpha_s)$ is Gell-Mann-Low function, b=11 for pure Yang-Mills QCD)
which vanishes in classical 
level. 
To obtain the mixing angle $\theta$, let us firstly choose operator $O$ to
be
 $O_\sigma=(\bar u u+\bar d d)/\sqrt{2}$. 
Assuming that
the lowest-lying  $0^{++}$ ground state, i.e. $\sigma$, saturates the
correlation function 
we obtain 
\begin{equation}
\displaystyle
{<|O_\sigma(0)|\sigma>\frac{1}{m_\sigma^2}<\sigma|{\cal O}|>=
-d_\sigma<|O_\sigma(0)|>}.
\label{s}
\end{equation}
Similarly, taking now the operator $O$ to be ${\cal O}$, we have 
\begin{equation}
\displaystyle
{<|{\cal O}(0)|H>\frac{1}{m_H^2}<H|{\cal O}|>=-d_G<|{\cal O}(0)|>}.
\label{g}
\end{equation}
In deriving (15), we have also assumed that
the ground state of $0^{++}$ "pure" glueball saturates the 
l.h.s of (\ref{g}). Defining $<|{\cal O}|H>=f_H$, 
$<|O_\sigma|\sigma>=f_\sigma$,
we can roughly define the effective field of the "pure" glueball H as  
\begin{equation}
\displaystyle{
H=\frac{1}{f_H}{\cal O}-\frac{1}{f_H}<|{\cal O}|>.
}.
\end{equation}
Therefore,
\begin{equation}
<\sigma|{\cal O}|>\approx \theta f_H . 
\end{equation}
From eqs (14-17), we get the mixing angle    
\begin{equation}
\displaystyle{
\theta=\frac{3m_\sigma^2}{f_\sigma f_H}<|O_\sigma|>=
\frac{3m_\sigma^2}{2f_\sigma m_H\sqrt{-<|{\cal O}|>}}<|O_\sigma|> } .
\label{theta}
\end{equation}   
Note that
the mixing angle $\theta$ is proportional to the quark's condense.
The parameter $f_\sigma$ in (\ref{theta}) can be estimated again by the QCD 
sum rule. Following\cite{scalar}, we have 
\begin{equation}
\begin{array}{ll}
f_\sigma^2=&\displaystyle{
\frac{6}{16\pi}M^4 \big (
1+\frac{13}{3}\frac{\alpha_s}{\pi}+
\frac{8\pi^2}{M^4}<|m\bar q q|>
}\\
&\displaystyle{
+\frac{\pi^2}{3M^4}<|\frac{\alpha_s}{\pi}
G^{\mu\nu}G_{\mu\nu}|>-\frac{1408}{81}\frac{\pi^3\alpha_s}{M^6}<|\bar q
q|>\big )e^{m_\sigma^2/M^2}},
\end{array}
\label{fs}     
\end{equation}
where $M$ is a parameter with mass dimension in Borel transformation,  
and $M \geq 1GeV$.
For the quark and gluon condensate, we take    
the standard values 
\begin{equation}
\begin{array}{l}
<|m\bar q q|>=(0.1Gev)^4,\\
<|\bar q q|>=<|\bar u u|>=<|\bar d d|>=-(0.25Gev)^3,\\
\displaystyle{
<|\frac{\alpha_s}{\pi}G^{\mu\nu}G_{\mu\nu}|>=(0.33Gev)^4}.\\
\end{array}
\end{equation} 
Taking a common value, $m_\sigma=500MeV$\cite{s1}
(\ref{theta}) gives 
\begin{equation}
\theta\approx 0.08 .
\label{va}
\end{equation}
The mixing angle $\theta$ in 
(\ref{va}) is consistent with that in (\ref{ss}).




In conclusion, we have studied the decay of the scalar glueball into
pions and its mixing 
 with the sigma particle in effective theory.
Our results show that $f_0(1500)$ decays into four pions through
2 $\sigma$ intermediate states. The $\rho$ can contribute to the
four-pion decay mode at one-loop level with branch ratio
\begin{equation}
\displaystyle {\frac{\Gamma(2\rho)}{\Gamma(2\sigma)}\sim 0.01}.
\end{equation}
Furthermore we fit our model to the recent experimental data on the decay of
$f_0(1500)$ and obtain that the mixing of glueball with sigma is small, which
 is consistent with our estimation based on low energy theorem
and QCD sum rule. 

{\bf ACKNOWLEDGMENTS}\\
We thank Zhenping Li, Qing Wang  and L.S. Kisslinger for discussions. This
work was supported in part by the national natural science foundation of
China.

\newpage

{\bf Figure Captions}\\
Fig.1 (a) Illustration of glueball decay into $2\sigma$ or $2\pi$;
(b)Illustration of glueball mixing with $\sigma$ induced by the
quark condense.\\
Fig.2 Ratio of $f_0(1500)$ decay into $2\sigma$ to $2\pi$ as function 
of $g_2$. In the numerical calculation, we take  $m_\sigma=500GeV$\\
Fig.3 Illustration of glueball mixing with $\sigma$ induced by the quark 
masses.

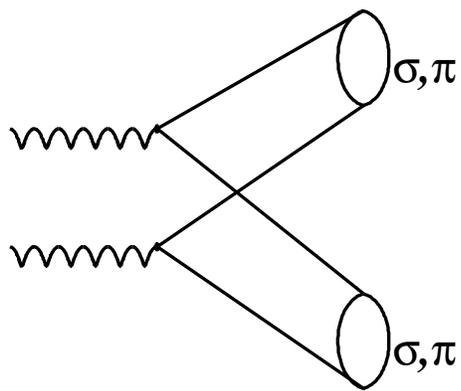
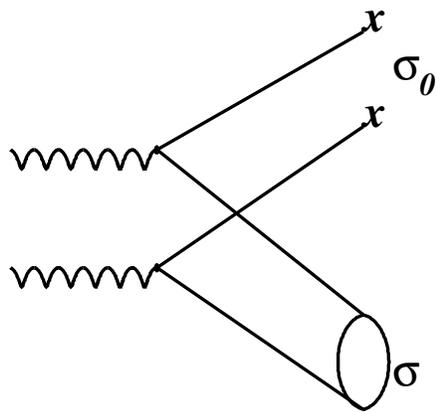
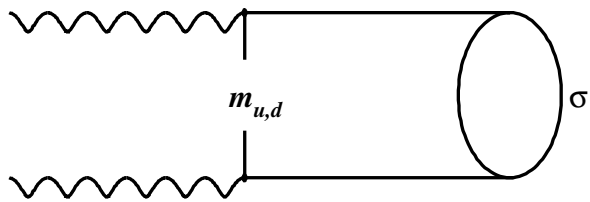
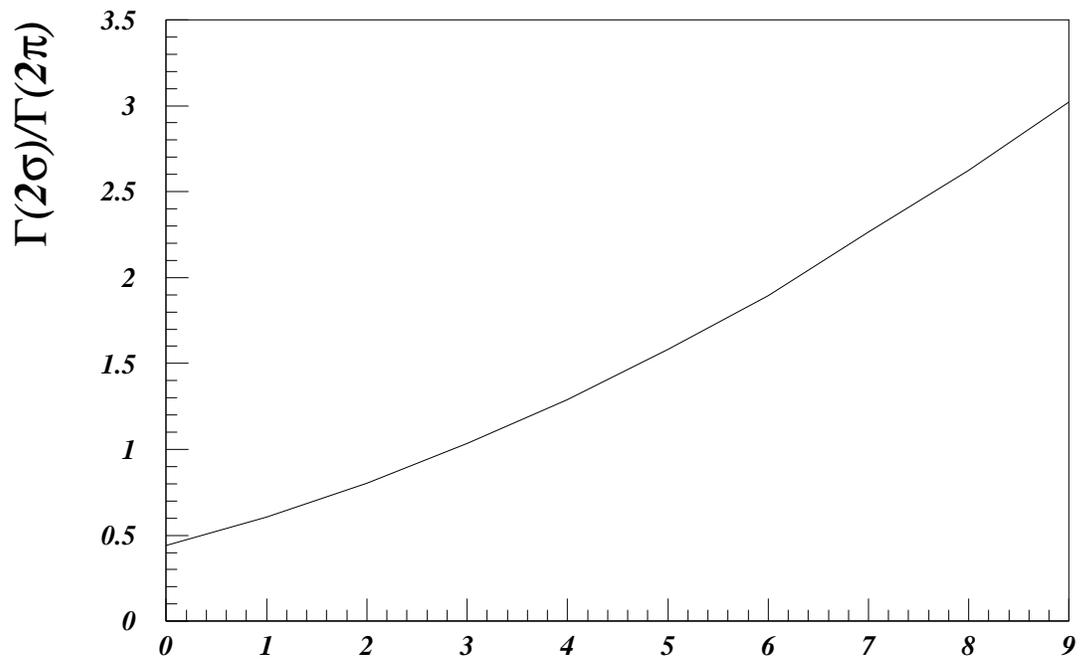
\end{document}